\documentclass[twocolumn,prl,showpacs,amsmath,amssymb,floatfix]{revtex4}

\usepackage{graphicx}% Include figure files
\usepackage{dcolumn}% Align table columns on decimal point
\usepackage{bm}% bold math

\begin{document}

\title{Cooperative Recombination of a Quantized High-Density
Electron-Hole Plasma}

\author{Y.~D.~Jho,$^{1,3}$ X.~Wang,$^{1}$ J.~Kono,$^{2}$
D.~H.~Reitze,$^{1}$ X.~Wei,$^{3}$ A.~A.~Belyanin,$^{4}$
V.~V.~Kocharovsky,$^{4,5}$ Vl.~V.~Kocharovsky,$^{5}$ and
G.~S.~Solomon$^{6}$}
\affiliation{$^1$Department of Physics,
University of Florida, Gainesville, Florida 32611, U.S.A.\\
$^2$Department of Electrical and Computer Engineering, Rice
University, Houston, Texas 77005, U.S.A.\\ $^3$National High
Magnetic Field Laboratory, Florida State University,
Tallahassee, Florida 32310, U.S.A.\\ $^4$Department of Physics,
Texas A\&M University, College Station, Texas 77843, U.S.A.\\
$^5$Institute of Applied Physics, Russian Academy of Sciences,
603950 Nizhny Novgorod, Russia\\ $^6$Solid-State Laboratories,
Stanford University, Stanford, California 94305, U.S.A.}

\date{\today}

\begin{abstract}
We investigate photoluminescence from a high-density electron-hole
plasma in semiconductor quantum wells created via intense
femtosecond excitation in a strong perpendicular magnetic field,
a fully-quantized and tunable system. At a critical magnetic
field strength and excitation fluence, we observe a clear
transition in the band-edge photoluminescence from
omnidirectional output to a randomly directed but highly
collimated beam. In addition, changes in the linewidth, carrier
density, and magnetic field scaling of the PL spectral features
correlate precisely with the onset of random directionality,
indicative of cooperative recombination from a high density
population of free carriers in a semiconductor environment.
\end{abstract}  \maketitle

%Introduction
The spontaneous appearance of macroscopic coherence is among the
most dramatic cooperative phenomena in condensed matter physics.
Superconductivity~\cite{DeGennes66} is the most prominent example,
and solid state analogs of Bose-Einstein condensation are actively
being pursued~\cite{Eisenstein04}. A related fundamental
cooperative process exists in quantum optics, superfluorescence
(SF)~\cite{Siegman, Zheleznyakov89, Bonifacio75}, in which an
incoherently prepared system of $N$ inverted atoms develops
macroscopic coherence self-consistently from vacuum fluctuations.
The resultant macroscopic dipole decays
superradiantly~\cite{Dicke54,Rehler71} producing a burst of
coherent radiation.  Superfluorescence has been observed in atomic
gases~\cite{Skribanowitz73,Gibbs77} and rarefied impurities in
crystals~\cite{Florian82,Zinoviev83,Malcuit87}.

The observation of cooperative recombination in semiconducting
systems is significantly more challenging due to ultrafast
decoherence.
%\textbf{ADDED BY DHR}
Nevertheless, investigations of such phenomena, especially in
strong magnetic fields, are important because they
allow us to probe quantum coherence in a previously inaccessible
regime. Coherent carrier dynamics in semiconductors has received
much attention in recent years owing to the development of
ultrafast lasers, but most of these investigations have focused
on excitons at relatively low densities~\cite{shah}.  Probing SF
emission in semiconductors provides insight into the cooperative
behavior
%[and coherent light emission]
of dense electron-hole ($e$-$h$) plasmas in quantum-engineered
semiconductors. Potentially, they can lead to novel sources of
intense, tunable, and ultrashort pulses in the near and
mid-infrared regions  and possibly establish new routes to
controlling electronic interactions at high densities using
strong magnetic fields.
%\textbf{END}

Pure SF~\cite{Siegman,Zheleznyakov89} is characterized by
several signatures that uniquely distinguish it from other
emission processes.  A SF pulse is bright, highly collimated, and
of short duration (less than the homogeneous dephasing time
$T_2$). SF is inherently random: polarization
fluctuations grow from initially incoherent quantum noise and
reach a macroscopic level. Thus, the emission direction varies
from shot to shot.  In addition, a SF pulse appears after a
delay time during which mutual coherence between
individual dipoles is established. Note the principal difference
between SF and superradiance
~\cite{Dicke54,Rehler71,Tokihiro93,Bjork94}: the latter develops
in a system in which the macroscopic polarization has been
initially established by an external laser field.
The key parameter governing the growth rate of cooperative
emission is the coupling strength between the electromagnetic
field and optical polarization, expressible as the cooperative
frequency, $\omega_c$~\cite{Zheleznyakov89,Belyanin}.  To
observe SF from a system of $e$-$h$ pairs in a
semiconductor quantum well (QW),
%%%
\begin{equation}
\omega_c = \sqrt{\frac{8 \pi^2 d^2 N \Gamma c}{\hbar n^2 \lambda
L_{QW}}} \label{coop}
\end{equation}
%%%
must exceed $2/T_2$ [or $2/(T_2 T_2^{*})^{1/2}$ if the
inhomogeneous dephasing time $T_2^{*} < T_2$]. Here $d$ is the
transition dipole moment; $N$, 2D $e$-$h$ density;
$\Gamma$, overlap of radiation with the QWs; $\hbar$, Planck's
constant; $n$, refractive index; $\lambda$,
wavelength (in vacuum); $c$, speed of light; and
$L_{\rm QW}$, total width of the QWs.
%\textbf{ADDED BY AB}
Under the optimal SF conditions, the pulse
duration scales as $\tau_{\rm SF} \sim 1/\omega_c \sim N^{-1/2}$
and thus the peak intensity scales as $I_{\rm SF} \sim \hbar
\omega N/\tau_{\rm SF} \sim N^{3/2} \sim
B^{3/2}$~\cite{Zheleznyakov89,Belyanin}; see also Sec. XII in
\cite{Bonifacio75}.
%\textbf{END}

Here, we investigate light emission in an undoped QW system in a
strong perpendicular magnetic field ($B$).  The field fully
quantizes the QW system into an atomic-like system with a series
of Landau levels (LLs) and thus strongly increases the density of
states (DOS) to accommodate a high-density $e$-$h$ plasma. We
measure emission as a function of laser fluence ($F_{\rm laser}$)
and $B$. The emission characteristics depend upon $N$ ($\sim
F_{\rm laser}$) and $B$, evolving from omnidirectional,
inhomogeneously broadened photoluminescence (PL) at low
densities and fields through a narrowly peaked but
omnidirectional output $\sim N$ to a randomly directed but
highly collimated output $\sim N^{3/2}\sim B^{3/2}$ as the field
increases.  The successive stages of emission progress from low
density PL through a regime where amplified spontaneous emission
(ASE) dominates into a regime characterized by stochastically
oriented but highly directional emission.

%Experiments
Samples were grown by molecular-beam epitaxy on GaAs,
consisting of a GaAs buffer followed by 15 layers of 8-nm
In$_{0.2}$Ga$_{0.8}$As QWs separated by 15-nm GaAs barriers and a
10-nm GaAs cap layer.  We used a 150~fs, 775~nm Ti:Sapphire
regenerative amplifier system, and measured
the PL as a function of $F_{\rm laser}$, $B$, and pump
spot size and geometry, up to 9.7~mJ/cm$^2$, 25~T, and
3$\times$3~mm$^2$ (the sample size), respectively.
Approximately 10$^{12}$ $e$-$h$ pairs/cm$^2$ were generated
for $F_{\rm laser}$ $\sim$ 0.01~mJ/cm$^2$. The laser beam was
delivered into a 31~T Bitter-type magnet through free space and
focused using a spherical or cylindrical lens. The QW
plane was perpendicular to $B$.  Emission was collected using
optical fibers from the opposite face and cleaved edges of the
sample and analyzed with a grating spectrometer equipped with a
charge-coupled device detector. Unless noted, each spectrum
consisted of the emission from $\sim$1000 pulses.
Two right-angle micro-prisms redirected the edge emission from
the sample to collection fibers. The collection area of the
prisms was 1$\times$1~mm$^2$, and the computed acceptance
angle based on geometry was $\sim$40$^{\circ}$.  Excitation at
775~nm creates carriers high in the
bands with an excess energy of 270~meV (the energy difference
between the initial carrier states and the 0-0
LL~\cite{MacDonald}), and thus a very short $T_2$ (a few fs).
The carriers then thermalize, becoming quantized with a longer
$T_2$. Thus, the resulting $e$-$h$ plasma is initially
\emph{completely incoherent}.

\begin{figure}
\includegraphics[scale=0.43]{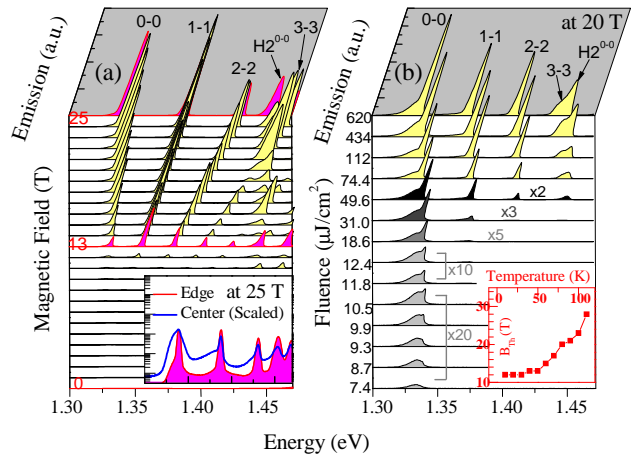}
\caption{(Color Online) Emission spectra as a function of (a)
magnetic field for fixed fluence (0.62~mJ/cm$^2$)
and (b) fluence for a fixed field (20~T).  To compare
the edge (red line) to center (blue line) collection in the
inset of Fig.~1(a), the opposite face is multiplied by 5000.
The inset in (b) shows how the threshold magnetic field depends
on temperature. \label{plspectra}} \end{figure}

%Results
Figure \ref{plspectra} shows spectra collected at the sample
edge perpendicular to the excitation direction, i.e., emission
in the QW plane (a) versus $B$ at a fixed $F_{\rm laser}$
and (b) versus $F_{\rm laser}$ at a fixed $B$ (20~T) at 10~K for
an excitation spot size of 0.5~mm. A threshold is observed in
both cases; inhomogeneously broadened PL peaks ($\sim$9 meV)
are seen at each interband LL transition until $F_{\rm laser}$
and $B$ exceed a threshold value, whereupon a narrow peak
($\sim$2 meV) emerges from the high-energy side of the broad
feature and dominates at high $F_{\rm laser}$. Identical spectra
are seen when collecting from above the pump spot, i.e.,
out-of-plane (left inset), although at a much lower efficiency.
Increasing or decreasing the pump spot size resulted in
qualitatively similar spectra for a given fluence.
%; broad PL was observed at low fluence or field while a narrow
%blue-shifted peak observed at higher fluence with much higher
%signal strength.
Thus, the observed behavior is not due to a spatially or
spectrally inhomogeneous distribution of carriers and indicates
the onset of stimulated emission.  Sharp emission features are
observed to 130~K (right inset), with the threshold field
for 0-0 LL increasing from 12~T to 28~T as the temperature
increases from 10~K to 110~K.

\begin{figure}
\includegraphics[scale=0.4] {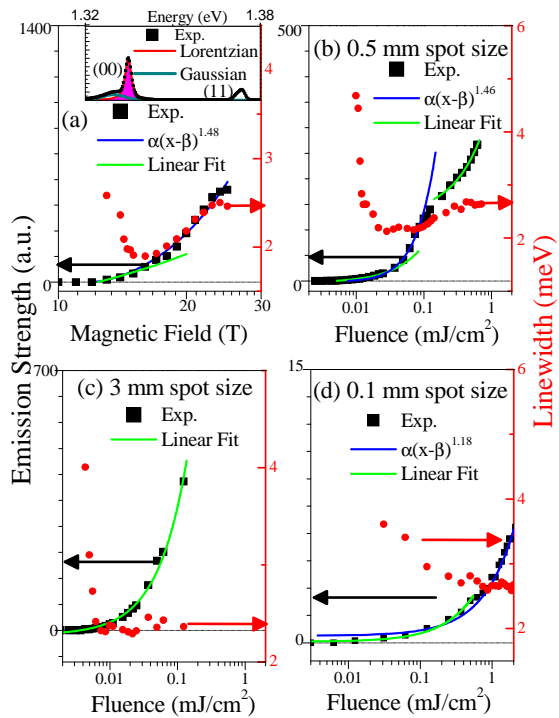}
\caption{(Color Online) Emission strength and linewidth of the
narrow peak from the 0-0 LL versus (a) $B$ and (b,c,d)
$F_{\rm laser}$ for different pump spot sizes. Both $B$ and
$F_{\rm laser}$ are on a logarithmic scale. The inset of
(a) shows the convolution method using a
Lorentzian for the sharp peak and a Gaussian for the broader
lower-energy peak.  The blue (green) lines in (a-d)
indicate the superlinear (linear) regime. \label{fluence}}
\end{figure}

The integrated strength of the 0-0 LL emission versus $B$
[Fig.~\ref{fluence}(a)] and $F_{\rm laser}$
[Fig.~\ref{fluence}(b)] was obtained from a Lorentzian lineshape
analysis of the narrow blueshifted feature for a
0.5~mm excitation spot. Below 12~T (and 0.01~mJ/cm$^2$), narrow
emission is not observed. In the range 12-14~T
(0.01-0.03~mJ/cm$^2$), the signal grows linearly (green lines)
with both $B$ and $F_{\rm laser}$. Above 14~T (0.03~mJ/cm$^2$),
the emission strength becomes superlinear (blue lines) with the
integrated signal $S \propto B^{3/2}$.
%\textbf{ADDED BY DHR}
Above 0.2~mJ/cm$^2$ in Fig.~\ref{fluence}(b), the signal resumes
a linear scaling.
%\textbf{END}
The linewidths (red circles), also plotted in
Figs.~\ref{fluence}(a) and \ref{fluence}(b) reveal a remarkable
correlation with the emission strength. In the linear regime,
the linewidth \emph{decreases} monotonically both versus $B$ and
$F_{\rm laser}$ until the emission becomes superlinear, where the
linewidth begins to \emph{increase}.  When the laser spot was
increased (decreased) to 3~mm (0.1~mm) as shown in
Figs.~\ref{fluence}(c) and \ref{fluence}(d), narrow emission was
observed, but both the integrated signal $S$ and the linewidth
exhibited qualitatively different scaling, and in both of these
cases, the linewidth monotonically decreased with increasing
fluence.

%\begin{figure}[tbp]
%\includegraphics[scale=0.3,trim=30 30 30 0,angle=-90] {fig3.ps}
%\caption{(Color Online) Edge emission spectra measured from two
%orthogonally aligned fibers for the angle $\theta$ at (a)
%0$^{\circ}$  and (b) 90$^{\circ}$, where $\theta$ is the angular
%separation between the longer beam axis and the direction of edge
%2 fiber as shown in the inset figure of (c). In (c), the emission
%strength of 0-0 LL is plotted for edge 1 (black) and edge 2 (red)
%as a function of angle. \label{cylind}}
%\end{figure}

\begin{figure}
\includegraphics[scale=0.44]{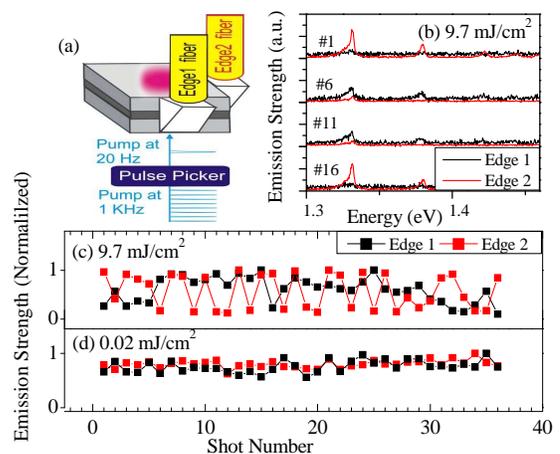} \caption{(Color
Online) (a) Experimental schematic showing single-shot
excitation and collection.  (b) Four representative emission
spectra from edge 1 (black) and edge 2 (red) fibers, excited
from single laser pulse and measured simultaneously. Normalized
emission strength from the 0-th LL versus shot number in the (c)
SF regime and (d) ASE regime. \label{directionality}} \end{figure}

Figure \ref{directionality} presents the directionality of the
emission for single pulse excitation.  Figure
\ref{directionality}(b) illustrates a series of spectra upon
\emph{single} pulse excitation at a fluence corresponding to the
superlinear emission regime in Fig.~\ref{plspectra}
(9.7~mJ/cm$^2$, 25 T) for a 0.5~mm diameter spot size collected
through edge 1 (black) and edge 2 (red). Figure
\ref{directionality}(c) displays the maximum peak height from
each edge (normalized to 1.0) versus shot number for the pumping
conditions in Fig.~\ref{directionality}(b). The maximum observed
emission strength in Fig.~\ref{directionality}(c) fluctuates as
much as eight times the minimum value, \emph{far greater} than
the pump pulse
fluctuation ($\sim$2\%). This strong anticorrelation between
signals received by different edges indicates \emph{a collimated
but randomly changing emission direction from pulse to pulse.} At
a lower $F_{\rm laser}$~=~0.02 mJ/cm$^2$ (obtained with a 3~mm
spot), qualitatively different behavior is seen;
Fig.~\ref{directionality}(d) shows omnidirectional emission on
every shot, as expected for ASE or PL.  We also probed the
spatial and directional characteristics of the emission process.
Using a cylindrical lens to generate a rod-like excitation
region, we measured the signal as a function of angle from
0$^{\circ}$ to 180$^{\circ}$ for $F_{\rm laser}$~=~0.02~mJ/cm$^2$
and $B$~=~25~T (not shown). The emission was highly directional
with a 40$^{\circ}$ divergence, comparable to the measurement
acceptance angle. In addition,
%the increased signal of $\sim$20
%along the long axis relative to the short axis is due to
%exponential gain, as expected for a stimulated process.
the measured amplitude ratio along the long axis relative to the
short axis was consistent with exponential gain, as expected for
a stimulated process.

The scaling of the LL peaks, the linewidth evolution, and emission
directionality point to the following evolution as $F_{laser}$ and
$B$ are increased: (i) In the low-density limit ($B$ $<$ 12~T,
$F_{\rm laser}$ $<$ 5 $\mu$J/cm$^2$), excited $e$-$h$ pairs relax
and radiate spontaneously through interband recombination. The
emission is isotropic with an inhomogeneous Gaussian width of
$\sim$9~meV. (ii) At a critical fluence $\sim$0.01~mJ/cm$^2$ and
$B$ = 12~T, population inversion is established and ASE develops,
leading to the emission of pulses. Figure \ref{directionality}(d)
shows that ASE is simultaneously emitted in all directions in the
plane. The reduction in linewidth with increasing fluence results
from conventional gain narrowing: spectral components near the
maximum of the gain spectrum are preferentially amplified than
components with greater detuning [Fig.~\ref{fluence}(a),(b)
below 17~T and 0.03~mJ/cm$^2$]. In the high gain regime, the
spectral width reduces to 2~meV FWHM but is still larger than
$2/T_2$. (iii) When the DOS and physical density are
sufficiently high, the cooperative frequency $\omega_c$
[Eq.~(\ref{coop})] exceeds $2/(T_2 T_2^{*})^{1/2}$. The $e$-$h$
pairs establish a macroscopic dipole after a short delay time
and emit a pulse through cooperative recombination (or a
sequence of pulses, depending on the pump fluence and the size
of the pumped area). Due to inhomogeneous broadening, the pulse
spectral width (inverse duration) can be smaller than the total
width of the radiation, so there is no strong broadening of the
line, as shown in Fig.~\ref{fluence}(a). Only at very high pump
powers the line broadens due to the reduced pulse duration,
until eventually saturation (due to the filling of all available
states) halts the further decrease in pulse duration. The
transition from ASE to cooperative emission at 0.03~mJ/cm$^2$ is
consistent with this observation. Significantly, we find that
unlike ASE, which should be emitted in all directions with the
same intensity [Fig.~\ref{directionality}(d)], in this regime
the initial quantum fluctuations grow to a macroscopic level to
establish coherence and lead to strong directional fluctuations
from shot to shot [Fig.~\ref{directionality}(c)].
%\textbf{ADDED BY DHR}
The return to linear scaling above 0.1~mJ/cm$^2$ is a
combined result of absorption saturation of the pump and
saturation of SF emission.
%\textbf{END}

%\textbf{ADDED BY AB AND MODIFIED BY DHR}
Since the data was collected in a time-integrated fashion, we
cannot directly probe the peak SF intensity scaling mentioned
after Eq.~(1). There are two lines of evidence indicating that
the observed superlinear scaling
% of the time-integrated signal intensities
%in Figs.~\ref{fluence}(a) and \ref{fluence}(b)
is due to the formation of {\em multiple} SF pulses from the 0-0
LL. The superlinear increase for the 0-0 emission is accompanied
by an emission decrease from higher LLs, implying a fast
depletion of the 0-0 level through SF followed by a
rapid relaxation of $e$-$h$ pairs from higher LLs and subsequent
re-emission. Also, the single pulse data shows that the two fiber
outputs are either correlated or anti-correlated in roughly equal
proportion. Fast relaxation from higher LLs replenishes the 0-0
LL, resulting in a second pulse of SF
emission in a random direction. On average, in 50\% shots both edges
receive a SF pulse, and in the other 50\% shots, only one edge
will receive both pulses, in qualitative agreement with
observations.
%\textbf{END}

It could be argued that the observed emission characteristics are
consistent with pure ASE (`lasing'), but this can be ruled out
as follows.
%There are multiple lines of evidence that rule out
%this interpretation.
Collimated, randomly directed emission, and superlinear scaling
are observed \emph{only} when the pumped spot is 0.5~mm,
approximately equal to the theoretically predicted coherence
length for SF emission in QWs, $L_c \sim c \tau_{\rm SF}
\ln(I_{\rm SF}/I_{\rm SE})$. They are not observed for 0.1~mm and
3~mm spot sizes. Furthermore, feedback from facet reflection is
naturally suppressed in our structures: propagation modeling
indicates that the emission is guided in the pumped region due
to polariton dispersion.
%\textbf{ADDED BY AB}
Optically excited $e$-$h$ pairs create a Lorentzian-type
dispersion of the refractive index $n \sim N \Delta/(\Delta^2 +
\gamma^2)$ near the central frequency of each interband LL
transition, where $\Delta$ is the detuning from the transition
frequency and $\gamma$ is the linewidth. The enhancement of $n$
on the high energy side of the inverted interband LL transition
in the pumped region, added to the background index contrast
between the MQW and substrate, is able to support
guided modes confined only in the pumped region. Such modes will
be blueshifted by $\Delta \sim \gamma$ with respect to the peak
of the spontaneous PL line, in agreement with
(Fig.~\ref{plspectra}).
%\textbf{END}
Once the radiation leaves the pumped region, it diverges such that
$\sim 5 \times 10^{-4}$ gets coupled back into the guided mode.
Self-guiding in the pumped region is essential for achieving
high-gain ASE and then SF. Additional suppression comes from the
area outside the pump region, since it acts as a high loss
absorber comprised an ensemble of two-level systems in the ground
state. Finally, pure SF does not require a rod-like geometry. As
shown in Refs.~\onlinecite{Zheleznyakov89,Lvovsky99}, cooperative
recombination is not constrained by the geometry of the excitation
region. Omnidirectional superfluorescent emission has been
observed in cesium~\cite{Lvovsky99}. Moreover, the disk-like
geometry of the pumped active region allows us to observe the key
evidence for SF, namely strong shot-to-shot fluctuations in the
emission direction. Previous experiments almost exclusively
employed a rod-like geometry, in which the only direct signature
of SF is the macroscopic fluctuations of the delay time of the SF
pulse. In a semiconductor system they would be manifested on the
sub-ps scale and very hard to observe. Finally, we note that prior
four wave mixing experiments on intra-LL
relaxation~\cite{Fromer2002} have shown that coherence is lost
within $<$ 250~fs, but these experiments were conducted at
much lower densities ($\sim$10$^{10}$ cm$^{-2}$) and magnetic
fields (8~T).  Densities (fields) in excess of 10$^{12}$
cm$^{-2}$ (12~T) are required for cooperative emission.

In conclusion, we have observed cooperative emission in a
strongly-coupled semiconductor system.  Using intense ultrafast
excitation and strong magnetic fields, the resulting density and
energy confinement is sufficient to generate a spontaneous
macroscopic polarization that decays through the emission of SF
pulses.  Our experiments observe this phenomenon by exploiting its
quantum-stochastic nature and demonstrate that photon-mediated
quantum coherence can develop spontaneously even in
strongly-interacting semiconductor systems.

We acknowledge support from the NSF
(DMR-0325474, ECS-0547019, and ECS-05011537) and the
NHMFL In-house Science Program. A portion of this
work was performed at the National High Magnetic Field
Laboratory, supported by NSF Cooperative Agreement
No.~DMR-0084173 and by the State of Florida.

%-------------------------------------------------------------------------

\end{document}